\documentclass[a4paper,12pt]{article}
\usepackage{epsfig}
\usepackage[utf8]{inputenc}
\usepackage{amsmath}
\usepackage{amsfonts}
\usepackage{multirow}
\usepackage{slashed}
\usepackage{amssymb}

\usepackage{array}
\usepackage{ragged2e}
\usepackage{float}
\usepackage{graphicx}
\graphicspath{ {Images/} }
\graphicspath{ {feyn/} }
\usepackage{xparse}
\usepackage{amsmath, amssymb, graphics, setspace}
\usepackage{rotating}
\usepackage{float}
\usepackage{epstopdf}
\RequirePackage{luatex85}
\usepackage{tikz}
\usepackage{amsmath}
\usepackage{mathtools}
\usepackage{amsfonts}
\usepackage{amssymb}
\usepackage{makeidx}
\usepackage{graphicx}
\usepackage{hyperref}
\usepackage{subfig}
\begin{document}

	\begin{center}
	\baselineskip 20pt 
	{\Large\bf 	Nonminimal Inflation in Supersymmetric 
		\\ GUTs with $U(1)_R \times Z_n$ Symmetry}
	
	\vspace{1cm}
	
	{\large 
	
		Muhammad Atif Masoud$^{a,}$\footnote{E-Mail: atifmasood23@gmail.com}, Mansoor Ur Rehman$^{a,}$\footnote{ E-Mail: mansoor@qau.edu.pk}, Mian Muhammad Azeem Abid	$^{a,}$\footnote{ E-mail:azeem$\_92$@live.com}
	} 
	
	\vspace{.5cm}

	{\baselineskip 20pt \it
		$^a$Department of Physics, Quaid-i-Azam University , \\ 
		Islamabad 45320, Pakistan \\
		\vspace{2mm} 
	}
	
	\vspace{1cm}
\end{center}
\begin{abstract}
	A supersymmetric hybrid inflation framework is employed to realize a class of non-minimal inflation models with $U(1)_R \times Z_n$ global symmetry. This framework naturally incorporates models based on grand unified theories by avoiding the most commonly faced monopole problem. The predictions of inflationary observables, the scalar spectral index $n_s = 0.960-0.966$ and the tensor to scalar ratio $r=0.0031-0.0045$, are in perfect agreement with the Planck 2018 data. For sub-Planckian values of the field the $Z_n$ symmetry is only allowed for $n\leq 4$.
\end{abstract}

\section{Introduction}
  One of the most favored inflationary model according to Planck 2018 results \cite{Akrami:2018odb} is the Starobinsky model \cite{Starobinsky:1980te}. The scalar field version of this model is equivalent to an inflationary model which exploits a strong non-minimal coupling of the scalar field with gravity. See for example \cite{Bezrukov:2008ej,Okada:2010jf} for a few of the non-supersymmetric models of non-minimal Higgs inflation. In order to realize non-minimal inflation in supersymmetric framework a special form of K\"ahler potential is employed. For the feasibility of realizing inflation with standard model like Higgs boson in the minimal supersymmetric standard model see \cite{Einhorn:2009bh,Ferrara:2010yw,Lee:2010hj}. Further this idea has also been applied to Higgs fields in grand unified theories (GUTs) \cite{Pallis:2011gr,Ahmed:2018jlv}.

   The supersymmetric hybrid inflation model provides an elegant framework to incorporate GUTs \cite{Dvali:1994ms}\cite{Copeland:1994vg}\cite{Linde:1997sj}. However, the standard version of supersymmetric hybrid inflation is plagued with the monopole problem which is a generic prediction of GUTs based on a simple gauge group. In this paper we effectively consider a model of non-minimal GUT Higgs inflation with a special form of K\"ahler potential that is usually employed in no-scale supergravity models \cite{Ellis:2013xoa}. In this model monopoles are produced during inflation and are inflated away. The viability of non-minimal inflation is explored in a broader context with an additional $Z_n$ symmetry. The predictions of various inflationary parameters are obtained in a generic GUT framework and are consistent with the Planck 2018 results.

\section{Superpotential with $U(1)_R \times Z_n$ Symmetry}
In a typical supersymmetric hybrid inflation framework based on a given GUT gauge group, $G$, we usually consider a gauge singlet superfield, $S$, along with a gauge non-singlet conjugate pair of Higgs superfields $H$ and $\overline H$.   Some of the examples of the GUT gauge groups are $SO(10)$, $SU(5)\times U(1)$ and $SU(4)_c \times SU(2)_L\times SU(2)_R$ with Higgs superfields residing in the $16, 10$ and $(4,1,2)$ dimensional representations of the respective gauge groups \cite{Senoguz:2003zw}. With this minimal content of superfields and the $U(1)_R \times Z_n$ global symmetry we obtain the following simple form of the superpotential \cite{Yamaguchi:2004tn,Senoguz:2004ky},
\begin{equation}\label{super}
W = \kappa S\left(-\mu^{2}+\frac{\left(H \overline{ H} \right)^{m}}{\Lambda^{2m-2}}\right).
\end{equation}
Here, $\kappa$ is a dimensionless coupling, $\mu$ is some superheavy mass and $\Lambda$ is the cut-off scale. Under $Z_n$ symmetry the superfield, $S$, carries zero charge whereas Higgs superfields carry unit charges. This makes the integer $m = n$ for odd values of $n$ and $m = n/2$ for even values of $n$ \cite{Senoguz:2004ky}. For example, values of $m=1,2,3$ correspond to $n=2,4,3$ respectively. However, for the special case of  $m=1$ we do not need to impose any $Z_2$ symmetry as GUT gauge symmetry alone is sufficient to restrict the form of the superpotential. Further, the superfield $S$ and the superpotential $W$ carry one unit of $R$ charge whereas $H \overline{ H}$ is neutral under $U(1)_R$ symmetry. This $R$ charge assignment ensures a linear relationship of $W$ in terms of $S$ which is necessary to realize a consistent model of inflation \cite{Dvali:1994ms}.

The global supersymmetric minimum occurs at
	\begin{equation}\label{eq:v2}
\left<S\right>=0, \,\,\, \left< \left( H \overline H \right)^m \right>  = M^{2m}  \equiv \mu^{2}\Lambda^{2m-2},
\end{equation}
where the Higgs vacuum expectation value (VEV) is described by $M$. This gauge symmetry breaking scale is taken to be the GUT scale, $M_{GUT} \equiv 2\times 10^{16}$ GeV, in our numerical calculations. Further we set $\Lambda = m_P$ where $m_P = 2 \times 10^{18}$ GeV is the reduced Planck mass.

The form of the superpotential considered above has been used before mostly in the context of new inflation in a supersymmetric framework. Once the field $S$ is stabilized we obtain an effective Higgs potential which, for values of fields below $M$, can be used for new inflation. For example, see \cite{Yamaguchi:2004tn} where it is used to realize pre-inflation in order to justify the initial conditions of new inflation. In ref.~\cite{Senoguz:2004ky}, it was used to realize a model of new inflation itself. In ref.~\cite{Antusch:2008gw}, flavon inflation is discussed using a similar form of the superpotential. For $SU(5)$ and flipped $SU(5)$ based GUT realization of new inflation see \cite{Rehman:2018gnr}. In this paper, however, we consider the other side of the Higgs potential where field values lie above $M$ and the potential is steep. A special form of the K\"{a}hler potential, which is usually employed in the no-scale gravity models, helps to reduce the slope of the potential and makes it suitable for the slow-roll conditions to apply. This setup gives rise to non-minimal Higgs inflation which is discussed below in detail with additional $Z_n$ symmetry. 
\section{Non-minimal Higgs Inflation with $Z_n$ Symmetry}
To achieve non-minimal inflation we consider the following special form of the K\"{a}hler potential
\begin{equation}\label{eq:v2}
\begin{split}
K&=-3m_{P}^{2}\log\Bigg(1-\frac{(|S|^{2}+|H|^{2}+ |\overline{ H}|^{2})}{3m_{P}^{2}}+\frac{\chi}{2m_{P}^{2}} \Big( \frac{(H\overline{ H})^{m}}{\Lambda^{2m-2} } + h.c \Big)  + \gamma \frac{|S|^{4}}{3m_{P}^{4}} \Bigg),
\end{split}
	\end{equation}
where $\chi$ and $\gamma$ are dimensionless parameters. This is a variant of the K\"{a}hler potential usually employed in the no-scale supergravity models where moduli fields are assumed to be stabilized \cite{Ellis:2013xoa}. The addition of last term is necessary for the stabilization of $S$ field \cite{Lee:2010hj}. The scalar potential and the metric in Jordan and Einstein frames are related via the conformal rescaling factor $\Omega^2 = e^{-K/3 m_{P}^{2}}$ as,
 \begin{equation}
 V_J = \Omega^{4} V_E, \quad  \quad  g_{J}^{\mu \nu} = \Omega^{2}  g_E^{\mu \nu}.
 \end{equation}
This defines the Einstein-frame scalar potential $V_E$ in terms of $W$ and $K$ as
\begin{equation}
V_{E} = e^{K/m_{P}^{2}}\left(K_{ij}^{-1}D_{z_{i}}WD_{z_{j}^{*}}W^{*} -3m_{P}^{-2}|W|^{2}\right) + V_D^E,
\end{equation}
where
\begin{eqnarray}
D_{z_{i}}W = \frac{\partial W}{\partial z_{i}}+\frac{1}{m_{P}^{2}}\frac{\partial K}{\partial z_{i}}W,
\quad K_{ij} = \frac{\partial^{2}K}{\partial z_{i} \partial z_{j}^{*}},\quad
D_{z_{j}^{*}}W^{*} = (D_{z_{i}}W)^{*},
\end{eqnarray}
with $z_{i}\in\{ S,H,\overline{H}\}$. Here, same notation has been used for the superfields and their scalar components. The Einstein-frame D-term potential is given by
\begin{equation}
V_{D}^E \propto  g^{2}  \, (|H|^{2}-|\overline H|^{2}).
\end{equation}
Writing complex Higgs fields in terms of real scalar fields,
\begin{eqnarray}\label{eq:v2}
H = \frac{\phi}{\sqrt{2}} \ e^{i \alpha} \cos \beta, \qquad \overline H  = \frac{\phi}{\sqrt{2}} \ e^{i \overline \alpha} \sin\beta,
\end{eqnarray}
the stabilized D-flat direction is obtained for $\beta = \pi/4$, $\alpha=\overline \alpha=0$ and this implies that
\begin{equation} 
H = \overline H = \frac{\phi}{2},
\end{equation} 
where $\phi$ is the canonically normalized real scalar field in the Jordan frame. 
Finally the scalar potential in the Einstein frame takes the following form
 \begin{equation}\label{eq:v2}
 V_{E} =\frac{\kappa^{2}\mu^{4}\left(1-\left(\frac{\phi}{2M}\right)^{2m}\right)^{2}}{\left(1-\frac{2}{3}\left(\frac{\phi}{2m_{P}}\right)^{2}+\chi \left(\frac{\phi}{2m_P}\right)^{2m}\right)^{2}}.
 \end{equation}
After conformal rescaling the canonically normalized inflaton field $\hat{\phi} (\phi)$ in the Einstein frame becomes a function of field $\phi$ as
\begin{equation}\label{eq:v2}
J(\phi) \equiv \left( \frac{d\hat \phi}{d\phi}\right)=\sqrt{\frac{1}{\Omega^2(\phi)}+\frac{3}{2}m_{P}^{2}\left(\frac{d \ln \Omega^2(\phi)}{d\phi}\right)^{2}}.
\end{equation}
The slow-roll parameters can now be expressed in terms of $\phi$ as
\begin{eqnarray}\label{eq:v2}
\epsilon(\phi) = \frac{1}{2}m_{P}^{2} \left(\frac{V_E^\prime}{J V_E}\right)^2,\quad \eta(\phi) = m_{P}^{2}\Bigg( \frac{V_E^{\prime \prime}}{J^2 V_E}- \frac{J^{\prime} V_E^\prime}{J^3 V_E}\Bigg) ,  
\end{eqnarray}
where a prime denotes a derivative with respect to $\phi$.  The scalar spectral index $n_s$ and the tensor to scalar ratio $r$ to the first order in slow-roll approximation are given by
\begin{equation}\label{eq:v2}
n_s\simeq 1-6 \epsilon(\phi_0) + 2 \eta(\phi_0), \qquad r \simeq 16 \epsilon(\phi_0),
\end{equation}
where the field value, $\phi_0$, corresponds to the number of e-folds, 
\begin{equation}\label{eq:n0}
N_0=\frac{1}{\sqrt{2}m_{P}}\int_{\phi_e}^{\phi_0}\dfrac{J(\phi)}{\sqrt{\epsilon(\phi)}}d\phi,
\end{equation}
before the end of inflation at $\phi = \phi_e$ defined by the condition $\epsilon(\phi_e)=1$. Also, $\phi_{0}$ corresponds to the pivot scale where the amplitude of the scalar power spectrum is normalized by Planck \cite{Akrami:2018odb} to be,
\begin{equation}\label{As}
A_{s}(k_{0})=\left.\frac{1}{24\,\pi^2 \epsilon(\phi)}\frac{V_E(\phi)}{m_P^4 }\right|_{\phi(k_{0})=\phi_0}= 2.137 \times 10^{-9},
\end{equation}
at $k_0=0.05\,\text{Mpc}^{-1}$. For non-minimal inflation with sub-Planckian values of the field we need to consider the large $\chi$ limit such that $\chi \left( \frac{\phi}{m_P}\right)^{2m}\gg 1$. Therefore, in the non-minimal limit with $\phi_0\gg M$, above relation can be used to eliminate $\kappa$ in terms of $\phi_0$ as
     	\begin{eqnarray}\label{eq:ratio}
     	\kappa \simeq \chi \sqrt{24 \pi^2 A_{s}(k_{0}) \epsilon(\phi_0)} \simeq \sqrt{32 \pi^2 A_{s}(k_{0})}\left(\frac{2m_P}{\phi_0}\right)^{2m}.
     	\end{eqnarray} 
Now we look for a relation of $\phi_0$ in terms of $N_0$. Using Eq.~(\ref{eq:n0}) and $\epsilon(\phi_e)=1$, the field values $\phi_{0}$ and $\phi_{e}$ can be written in terms of $N_{0}$ as
\begin{figure}[t]
	\includegraphics[width=6.8cm]{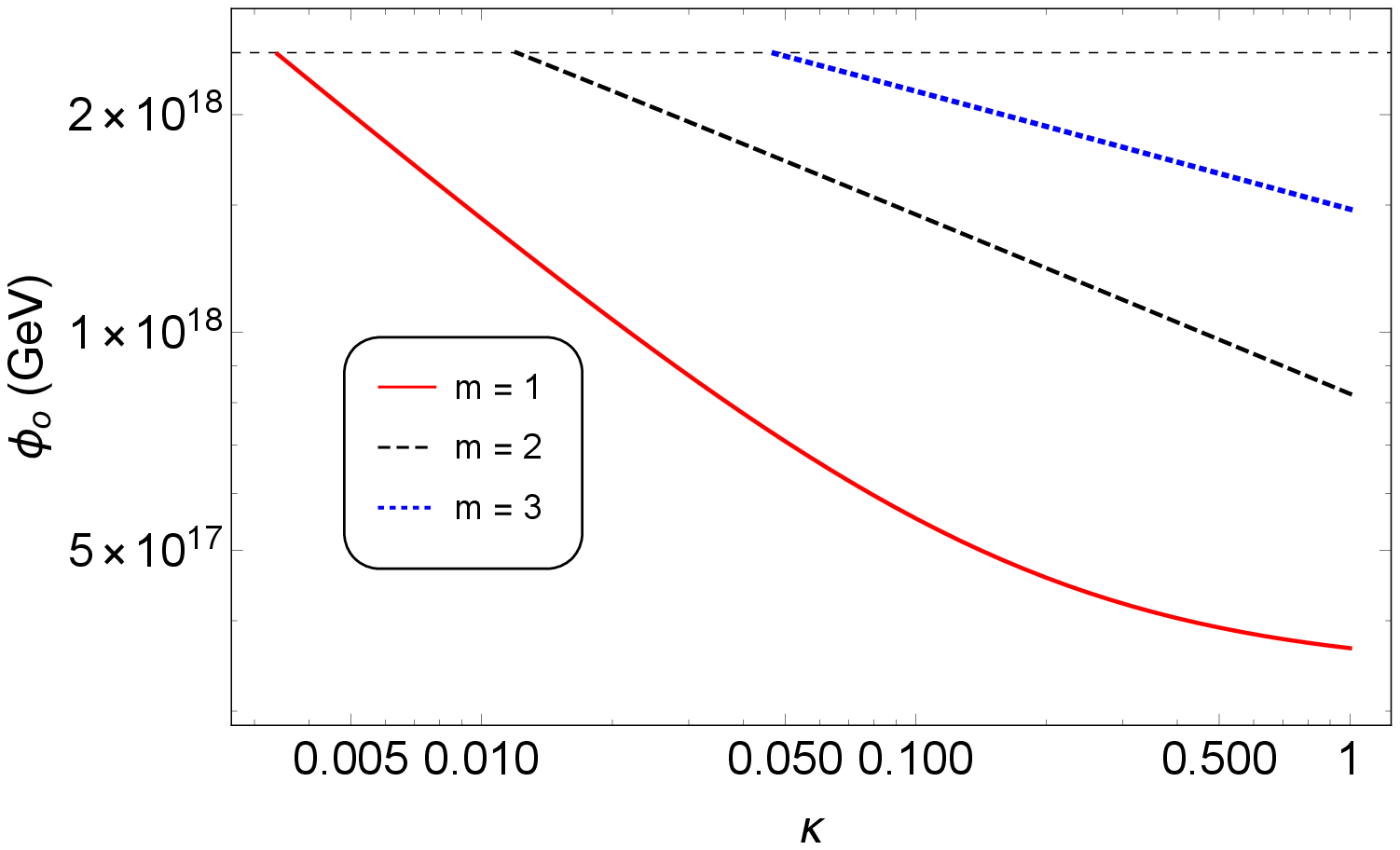} %
	\includegraphics[width=6.8cm]{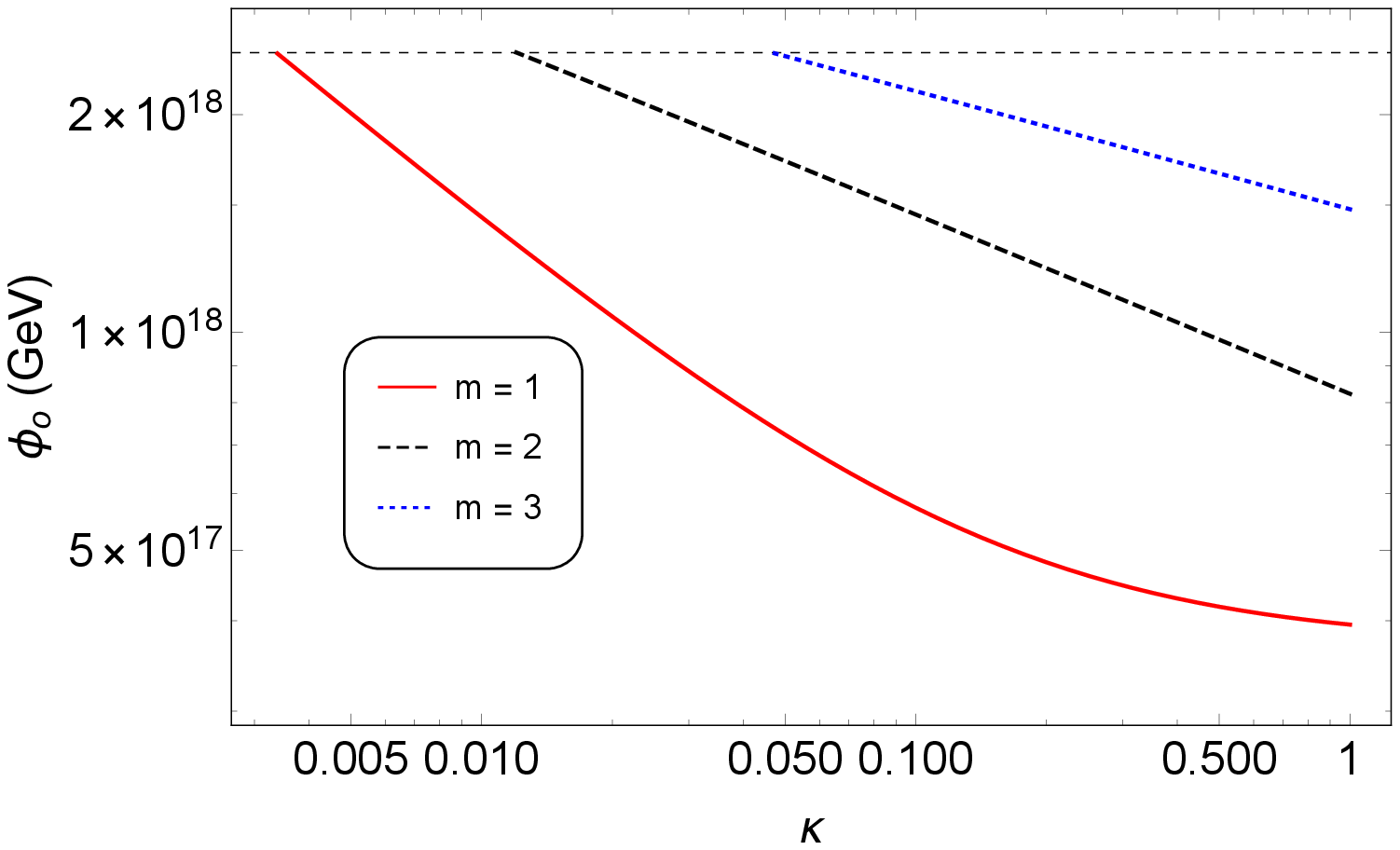} %
	\caption{The variation of field value $\phi_0$ versus $\kappa$ for $m=1,2,3$ and $N_{0} = 50$(left panel), $60$(right panel). We set the gauge symmetry breaking scale $M = 2 \times 10^{16}$ GeV.}
	\label{fig:phik}%
\end{figure}
     	 \begin{equation} \label{phi}
     	\frac{\phi_{0}}{2 m_P} \simeq \left( \frac{4N_0}{3 \chi} \right)^{1/2m}, \quad 
          \frac{\phi_{e}}{2 m_P} \simeq   \left( \frac{4}{3 \chi^2} \right)^{1/4m}.
     	 \end{equation}
Therefore, field values are expected to change with $m$ or $Z_n$ symmetry. This is confirmed by the exact numerical results shown in the Fig.~(\ref{fig:phik}) for the variation of $\phi_0$ with respect to $\kappa$. With sub-Planckian field values $\phi_0 \lesssim m_P$ we obtain $\chi \gtrsim \frac{2^{2m+2}}{3}N_0 \gg 1$ which provides a cross-check for using the large $\chi$ limit at first place. Using again above value of $\phi_0$ in Eq.~(\ref{eq:ratio}) we obtain a constant value for the ratio $\kappa / \chi$ written in terms of $N_0$ as
     	 \begin{figure}[t]
	\includegraphics[width=6.8cm]{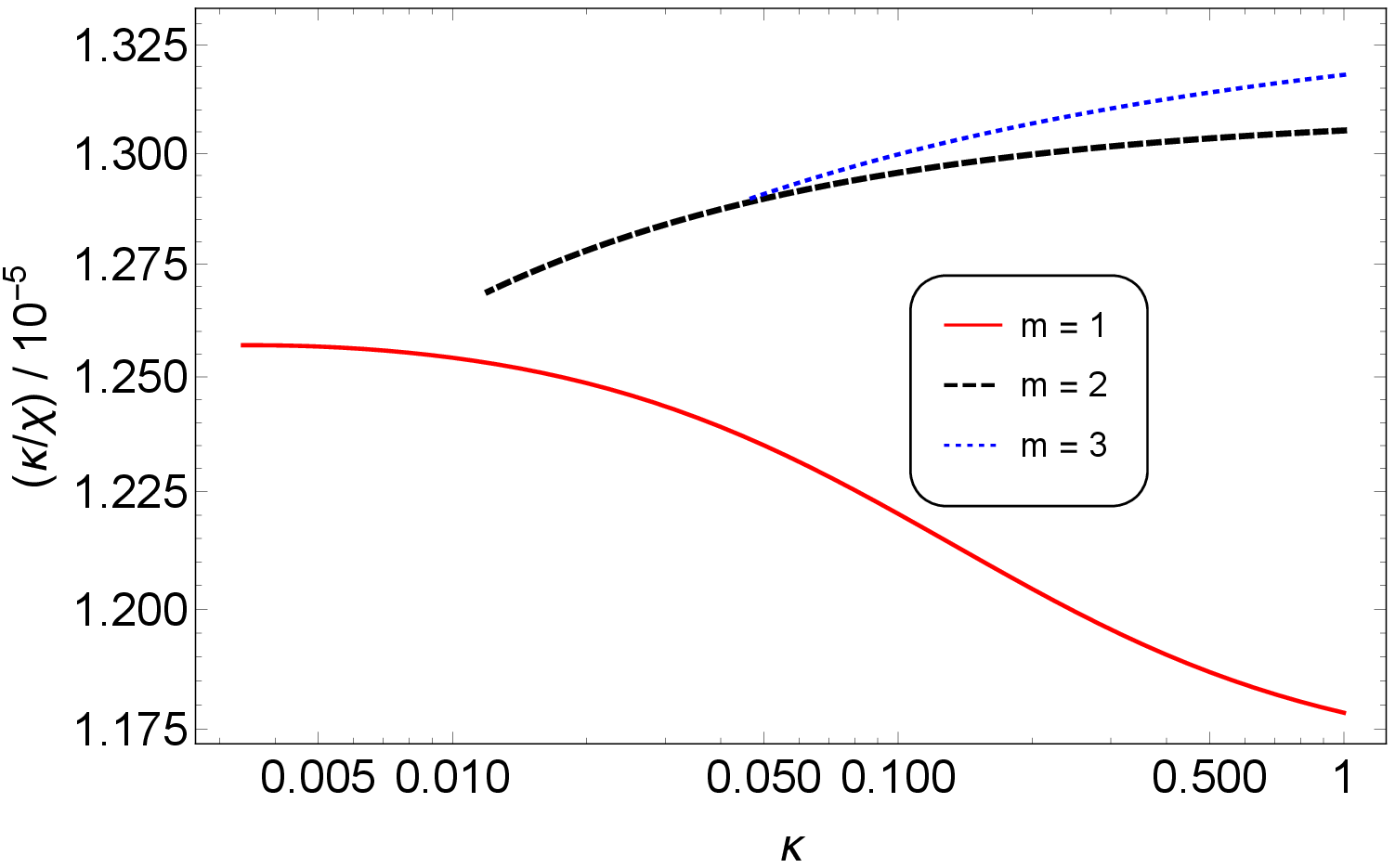} %
	\includegraphics[width=6.8cm]{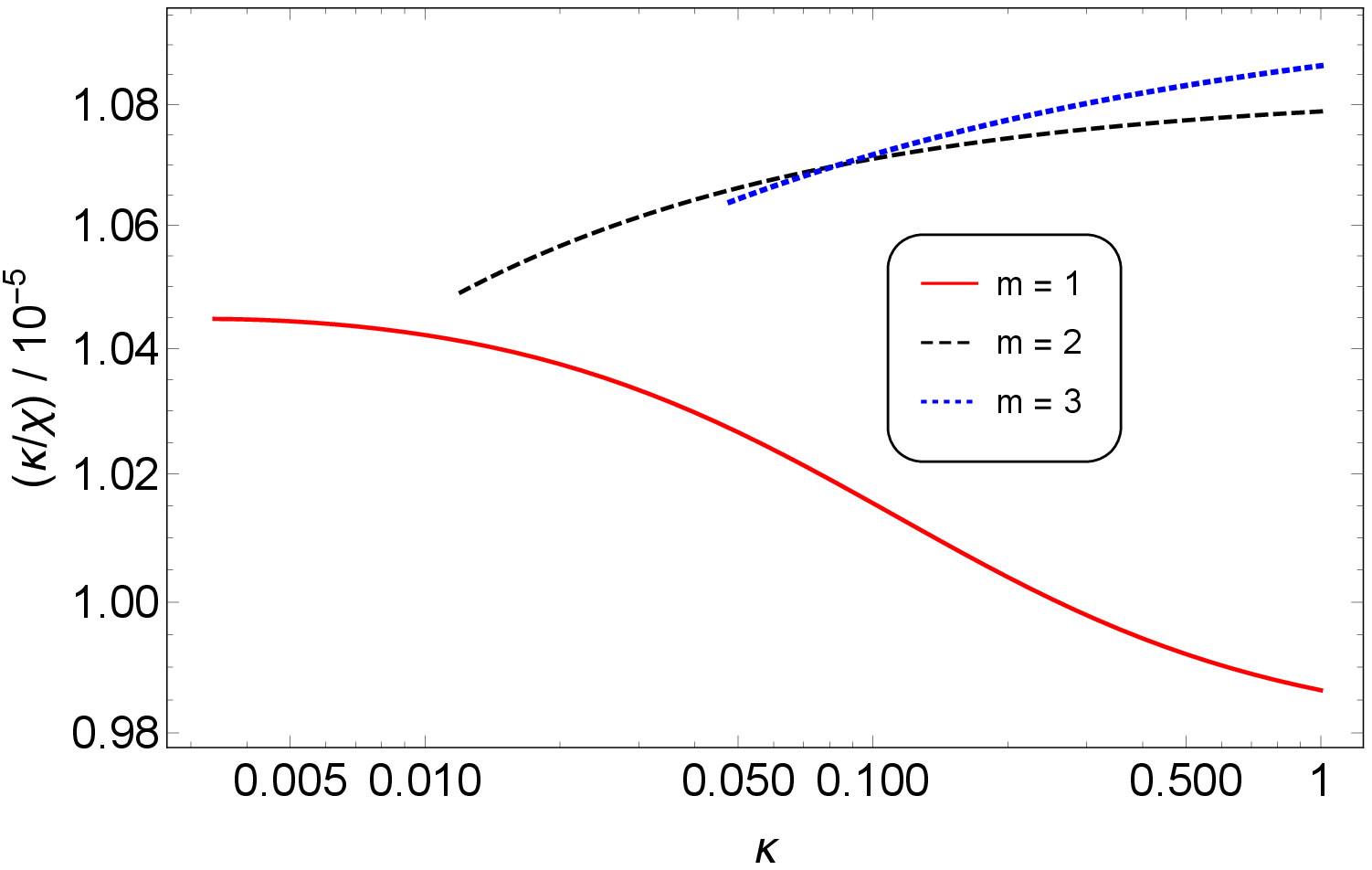} %
	\caption{The ratio $\kappa/\chi$ versus $\kappa$ for $m=1,2,3$ and $N_{0} = 50$ (left panel), $60$(right panel). We set the gauge symmetry breaking scale $M = 2 \times 10^{16}$ GeV.}	\label{fig:chik}%
\end{figure}

     	 \begin{equation} \label{ratio}
     	\kappa/\chi \simeq \frac{3\sqrt{2 \pi^2 A_{s}(k_{0})}}{N_0} \simeq \left\{ \begin{array}{ll}
         1.23 \times 10^{-5} & \mbox{for $ N_0 = 50,$} \\
        1.03 \times 10^{-5} & \mbox{for $ N_0 = 60$}.\end{array} \right.  
     	 \end{equation}
This ratio turns out to be of order $10^{-5}$ showing a weak dependence on $m$ or $Z_n$ symmetry in the non-minimal limit and this can also be seen in our numerical results displayed in the Fig.~(\ref{fig:chik}). 
We can express field value $\phi_0$ in terms of $\kappa$ using Eq.~(\ref{ratio}) in Eq.~(\ref{phi}) as
\begin{equation}
\frac{\phi_0}{m_P} \simeq  2 \left( \frac{4N_0/\kappa}{3 \times 10^{5}} \right)^{1/2m}.
\end{equation}
For a given value of $\kappa$, this expression explains the observed increasing trend of field values with respect to $m$ as shown in Fig.~\ref{fig:phik}. This trend leads to fine tuning in the solutions with large values of $m$ as soon as $\phi_0$ becomes tran-Planckian. Therefore, we allow $m \leq 3$ or $n \leq 4$ for $\phi_0 \lesssim m_P$ with perturbative values of $\kappa \lesssim 0.1$. For $SU(5)$ GUT with Higgs field in the adjoint representation we expect to obtain two more solutions for $m = 3/2$ and $ m = 5/2$ effectively.

Finally, the expression of $\phi_0$ is used to obtain the scalar spectral index $n_{s}$ and the tensor to scalar ratio $r$ in terms of $N_0$,
     	\begin{equation}\label{eq:v2}
     	n_{s} \simeq 1-\frac{2}{N_0} \simeq \left\{ \begin{array}{ll}
         0.960 & \mbox{for $ N_0 = 50,$} \\
        0.967 & \mbox{for $ N_0 = 60$},\end{array} \right.   \quad 
     	r \simeq \frac{12}{N_0^2} \simeq \left\{ \begin{array}{ll}
         0.0048 & \mbox{for $ N_0 = 50,$} \\
        0.0033 & \mbox{for $ N_0 = 60$},\end{array} \right. 
     	\end{equation}
where, $\epsilon (\phi_0) = \frac{3}{4N_0^2}$ and $\eta(\phi_0) = -\frac{1}{N_0}$. This result holds in the leading order approximation and also explains the weak dependence of $n_s$ and $r$ on $m$ or $Z_n$ symmetry as confirmed by our numerical estimates displayed in Table-I. We show the predictions of various inflationary parameters in Table-I, using first order slow-roll approximation, for $\kappa = 0.1$, $M = 2\times 10^{16}$ GeV and $N_0 = (50,60)$. We obtain $n_s \simeq 0.96\,(0.966)$ and $r \simeq 0.0045\,(0.0031)$ for $N_0 = 50\,(60)$ respectively, independent of $m$ and $\kappa$ values. The non-minimal coupling parameter is large $\chi \sim 10^5$ and this is a common feature of these models. An order of magnitude  estimate of error expectancy in inflationary parameters can be calculated from the second order slow-roll contribution. This can be described as a fractional change in the corresponding quantity, e.g., $\Delta n_s/n_s \simeq 0.01\%$, $\Delta r/r \simeq 1.5\%$, $\Delta \phi/\phi \simeq 0.4\%$, $\Delta \chi/\chi \simeq 1\%$. 
\begin{table}[t]
		\begin{center} 
\begin{tabular}{|c| c| c| c| c| c| c| c|}
		\hline
		\multicolumn{6}{|c|}{ For $N_{0}$ = 50} \\
		\hline\hline
		
	                                   	&  $r$       &    $n_{s}$     & $\phi_{0}$                  &    $\phi_{e}$                        &  $\chi$     \\
		\hline \hline	$m=1$           & 0.0045     &    0.960       &   $5.5 \times10^{17}$      &    $7.4\times10^{16}$               &  $8194$         \\
		
		\hline	    	$m=2$           & 0.0051     &    0.957      &  $1.4\times10^{18}$         &   $4.0\times10^{17}$                &  $7718$      \\
		
		\hline	        $m=3$           & 0.0051     &    0.957      &  $2.1 \times10^{18}$          &   $8.6\times10^{17}$                &  $7693$        \\
		\hline
		\multicolumn{6}{|c|}{For $N_{0}$ = 60} \\
		\hline\hline

		\hline \hline $m=1$              & 0.0031   &   0.966  &  $ 5.7\times10^{17}$      &    $7.2\times10^{16}$ &      $9848$          \\
		
		\hline        $m=2$              & 0.0035   &   0.965  &  $ 1.4\times10^{18}$      &    $3.8\times10^{17}$ &       $9337$            \\
		
		\hline	      $m=3$              & 0.0035   &   0.964  &  $ 2.2\times10^{18}$      &    $8.3\times10^{17}$ &      $9331$          \\
		\hline \hline
		\end{tabular}
		\caption{The predicted values of inflationary parameters with gauge symmetry breaking scale $M = 2 \times 10^{16}$GeV and $\kappa = 0.1$.} 
		\end{center} 
\end{table}

       \section{Conclusion}
We have studied a class of models based on the realization of non-minimal inflation in $R$-symmetric supersymmetric hybrid inflation framework with an additional $Z_n$ symmetry. The requirement of sub-Planckian field values is satisfied in the large $\chi$ limit. This also restricts the possible values of $n\leq4$ with $\kappa \lesssim 0.1$. We have calculated the predictions of $n_s$ and $r$ numerically and also provided the analytic justification of these results. Finally, we conclude that the results of non-minimal inflation hold in a rather broad class of supersymmetric GUT models.


\begin{thebibliography}{0}   

\bibitem{Akrami:2018odb} 
  Y.~Akrami {\it et al.} [Planck Collaboration],
  arXiv:1807.06211 [astro-ph.CO].


\bibitem{Starobinsky:1980te} 
  A.~A.~Starobinsky,
  Phys.\ Lett.\ B {\bf 91}, 99 (1980)
  [Phys.\ Lett.\  {\bf 91B}, 99 (1980)]
  [Adv.\ Ser.\ Astrophys.\ Cosmol.\  {\bf 3}, 130 (1987)].



\bibitem{Bezrukov:2008ej} 
  F.~L.~Bezrukov, A.~Magnin and M.~Shaposhnikov,
  Phys.\ Lett.\ B {\bf 675}, 88 (2009)
  doi:10.1016/j.physletb.2009.03.035
  [arXiv:0812.4950 [hep-ph]];
  A.~De Simone, M.~P.~Hertzberg and F.~Wilczek,
  Phys.\ Lett.\ B {\bf 678}, 1 (2009)
  doi:10.1016/j.physletb.2009.05.054
  [arXiv:0812.4946 [hep-ph]];
  A.~O.~Barvinsky, A.~Y.~Kamenshchik, C.~Kiefer, A.~A.~Starobinsky and C.~Steinwachs,
  JCAP {\bf 0912}, 003 (2009)
  doi:10.1088/1475-7516/2009/12/003
  [arXiv:0904.1698 [hep-ph]];
  N.~Okada, M.~U.~Rehman and Q.~Shafi,
  arXiv:0911.5073 [hep-ph].
  
\bibitem{Okada:2010jf} 
  N.~Okada, M.~U.~Rehman and Q.~Shafi,
  Phys.\ Rev.\ D {\bf 82}, 043502 (2010)
  doi:10.1103/PhysRevD.82.043502
  [arXiv:1005.5161 [hep-ph]];
  A.~Linde, M.~Noorbala and A.~Westphal,
  JCAP {\bf 1103}, 013 (2011)
  doi:10.1088/1475-7516/2011/03/013
  [arXiv:1101.2652 [hep-th]];
  N.~Okada, M.~U.~Rehman and Q.~Shafi,
  Phys.\ Lett.\ B {\bf 701}, 520 (2011)
  doi:10.1016/j.physletb.2011.06.044
  [arXiv:1102.4747 [hep-ph]];
  C.~Pallis and Q.~Shafi,
  JCAP {\bf 1503}, no. 03, 023 (2015)
  doi:10.1088/1475-7516/2015/03/023
  [arXiv:1412.3757 [hep-ph]];
  N.~Bostan, Ö.~Güleryüz and V.~N.~Şenoğuz,
  JCAP {\bf 1805}, no. 05, 046 (2018)
  doi:10.1088/1475-7516/2018/05/046
  [arXiv:1802.04160 [astro-ph.CO]];
  N.~Bostan and V.~N.~Şenoğuz,
  arXiv:1907.06215 [astro-ph.CO].
  

  
\bibitem{Einhorn:2009bh} 
  M.~B.~Einhorn and D.~R.~T.~Jones,
  JHEP {\bf 1003}, 026 (2010)
  doi:10.1007/JHEP03(2010)026
  [arXiv:0912.2718 [hep-ph]].
    
    
    \bibitem{Ferrara:2010yw} 
    S.~Ferrara, R.~Kallosh, A.~Linde, A.~Marrani and A.~Van Proeyen,
    Phys.\ Rev.\ D {\bf 82}, 045003 (2010)
    doi:10.1103/PhysRevD.82.045003
    [arXiv:1004.0712 [hep-th]].
  	 

  \bibitem{Lee:2010hj} 
  H.~M.~Lee,
  JCAP {\bf 1008}, 003 (2010)
  doi:10.1088/1475-7516/2010/08/003
  [arXiv:1005.2735 [hep-ph]].
    

    
    \bibitem{Pallis:2011gr} 
    C.~Pallis and N.~Toumbas,
    JCAP {\bf 1112}, 002 (2011)
    doi:10.1088/1475-7516/2011/12/002
    [arXiv:1108.1771 [hep-ph]].

        	      
       \bibitem{Ahmed:2018jlv}
       W.~Ahmed and A.~Karozas,
       Phys.\ Rev.\ D {\bf 98} (2018) no.2,  023538
       doi:10.1103/PhysRevD.98.023538
       [arXiv:1804.04822 [hep-ph]].
      
       
      
   \bibitem{Dvali:1994ms} 
   G.~R.~Dvali, Q.~Shafi and R.~K.~Schaefer,
   Phys.\ Rev.\ Lett.\  {\bf 73}, 1886 (1994)
   [hep-ph/9406319].
   \bibitem{Copeland:1994vg} 
   E.~J.~Copeland, A.~R.~Liddle, D.~H.~Lyth, E.~D.~Stewart and D.~Wands,
   Phys.\ Rev.\ D {\bf 49}, 6410 (1994)
   [astro-ph/9401011].
   \bibitem{Linde:1997sj} 
   A.~D.~Linde and A.~Riotto,
   Phys.\ Rev.\ D {\bf 56}, R1841 (1997)
   [hep-ph/9703209];
    	V.~N.~Senoguz and Q.~Shafi,
    	Phys.\ Rev.\ D {\bf 71}, 043514 (2005)
    	doi:10.1103/PhysRevD.71.043514
    	[hep-ph/0412102];
  M.~U.~Rehman, Q.~Shafi and J.~R.~Wickman,
  Phys.\ Lett.\ B {\bf 683}, 191 (2010)
  doi:10.1016/j.physletb.2009.12.010
  [arXiv:0908.3896 [hep-ph]].
  

\bibitem{Ellis:2013xoa} 
  J.~Ellis, D.~V.~Nanopoulos and K.~A.~Olive,
  Phys.\ Rev.\ Lett.\  {\bf 111}, 111301 (2013)
  Erratum: [Phys.\ Rev.\ Lett.\  {\bf 111}, no. 12, 129902 (2013)]
  doi:10.1103/PhysRevLett.111.129902, 10.1103/PhysRevLett.111.111301
  [arXiv:1305.1247 [hep-th]].
  
  
\bibitem{Senoguz:2003zw} 
  V.~N.~Senoguz and Q.~Shafi,
  Phys.\ Lett.\ B {\bf 567}, 79 (2003)
  doi:10.1016/j.physletb.2003.06.030
  [hep-ph/0305089].
  
    
\bibitem{Yamaguchi:2004tn} 
  M.~Yamaguchi and J.~Yokoyama,
  Phys.\ Rev.\ D {\bf 70}, 023513 (2004)
  doi:10.1103/PhysRevD.70.023513
  [hep-ph/0402282].
       

    	 \bibitem{Senoguz:2004ky} 
    	 V.~N.~Senoguz and Q.~Shafi,
    	 Phys.\ Lett.\ B {\bf 596}, 8 (2004)
    	 doi:10.1016/j.physletb.2004.05.077
    	 [hep-ph/0403294].
    	 


\bibitem{Antusch:2008gw} 
  S.~Antusch, S.~F.~King, M.~Malinsky, L.~Velasco-Sevilla and I.~Zavala,
  Phys.\ Lett.\ B {\bf 666}, 176 (2008)
  doi:10.1016/j.physletb.2008.07.051
  [arXiv:0805.0325 [hep-ph]].
  
    
 
       \bibitem{Rehman:2018gnr}
       M.~U.~Rehman, M.~M.~A.~Abid and A.~Ejaz,
       arXiv:1804.07619 [hep-ph].
       
        	    	 
    	
  
  
\end{thebibliography}
\end{document}